\documentclass[aps,twocolumn,floatfix,superscriptaddress]{revtex4-2}
\usepackage{amsfonts}
\usepackage{amssymb}
\usepackage{amsmath}

\usepackage{mathtools}
\usepackage{physics}
\usepackage{graphicx}
\usepackage{calc}
\usepackage{amsmath}
\usepackage{indentfirst}
\usepackage{dsfont}
\usepackage[T1]{fontenc}
\usepackage[utf8]{inputenc}
\usepackage[english]{babel}
\usepackage{braket}
\usepackage[normalem]{ulem}
\usepackage{svg}
\usepackage{dcolumn}
\usepackage{bm}
\usepackage{bbold}
\usepackage{soul}
\usepackage{hyperref}
 \graphicspath{{{new_plots}}}

\newcommand{\ak}[1]{\textcolor{blue}{#1}}

\def\be{\begin{equation}}
\def\ee{\end{equation}}
\def\bea{\begin{eqnarray}}
\def\eea{\end{eqnarray}}
\def\bi{\begin{itemize}}
\def\ei{\end{itemize}}
\def\bin{\begin{enumerate}}
\def\ein{\end{enumerate}}

\def\ra{\rangle}

\begin{document}

\title{Phonon-assisted coherent transport of excitations in Rydberg-dressed atom arrays}

\author{Arkadiusz Kosior}
\affiliation{Institute for Theoretical Physics, University of Innsbruck, 6020 Innsbruck, Austria}
\author{Servaas Kokkelmans}
\affiliation{Department of Applied Physics, Eindhoven University of Technology, PO Box 513, 5600 MB Eindhoven, The Netherlands}
\author{Maciej Lewenstein}
\affiliation{Institut de Ciencies Fotoniques, The Barcelona Institute of Science and Technology, 08860 Castelldefels, Barcelona, Spain}
\affiliation{ICREA, Passeig Lluis Companys 23, 08010 Barcelona, Spain}
\author{Jakub Zakrzewski}
\affiliation{Instytut Fizyki Teoretycznej, Uniwersytet Jagiello\'nski, \L{}ojasiewicza 11, 30-348 Krak\'ow, Poland}
\affiliation{Mark Kac Center for Complex Systems Research, Jagiellonian University, \L{}ojasiewicza 11, 30-348  Krak\'ow, Poland}
\author{Marcin P\l{}odzie\'n}
\affiliation{Institut de Ciencies Fotoniques, The Barcelona Institute of Science and Technology, 08860 Castelldefels, Barcelona, Spain}

\begin{abstract}
Polarons, which arise from the self-trapping interaction between electrons and lattice distortions in a solid, have been known and  extensively investigated for nearly a century. Nevertheless, the study of polarons continues to be an active and evolving field, with ongoing advancements in both fundamental understanding and practical applications. Here, we present a microscopic model that exhibits a diverse range of dynamic behavior, arising from the intricate interplay between two excitation-phonon coupling terms.
The derivation of the model is based on an experimentally feasible Rydberg-dressed system with dipole-dipole interactions, making it a promising candidate for realization in a Rydberg atoms quantum simulator  for excitation dynamics interacting with optical phonons.  Remarkably, our analysis reveals a growing asymmetry in Bloch oscillations, leading to a macroscopic transport of non-spreading excitations under a constant force.  Finally,
we demonstrate the robustness of our findings against on-site random potential.
\end{abstract}
\maketitle

\section{Introduction}

Polarons are quasi-particles that emerge from the coupling between electrons (or holes) with ions of a crystalline structure in polarizable materials. The idea of electron self-trapping due to lattice deformations dates back to Landau's seminal 1933 paper \cite{Landau}, but the modern concept of a polaron as an electron dressed by phonons was formulated in 1946 by Pekar \cite{Pekar1946}, and developed later by Fr\"ohlich \cite{Frhlich1954}, Feynman \cite{Feynman1955, Feynman62}, Holstein \cite{Holstein1959}, and Su, Schrieffer and Heeger~\cite{Su1979,Chao1985,Heeger1988}.  
Since their discovery, polarons have been extensively investigated, both theoretically and experimentally, not only in the field of condensed matter physics (for reviews see Refs.~\cite{Alexandrov,Franchini2021}), but also in various chemical and biological contexts, e.g., in protein propagation \cite{Chuev1993,evizovi2010,Georgiev2019}. In particular, in the modeling of charge migration in DNA molecules, it is assumed that a localized polaron is formed in the helix near a base due to an interaction between a charge carrier and a phonon. When a uniform electric field is applied, the polaron moves at a constant velocity, and a current flows through the chain \cite{Singh2004,Trugman2013,Fornari2014}. The charge carrier transport takes place due to coupling between carrier and phonons; in contrast, in the absence of phonons, an external constant force induces Bloch oscillations \cite{Glueck02,Wiater2017,PhysRevLett.126.103002}, where the mean position of the carrier is constant while its  width periodically changes in time. 

Polarons have been studied in many, seemingly different experimental setups, ranging from ultracold ions 
\cite{Solano2012,Cirac2012,Lamata2014,PhysRevResearch.2.033326}, polar molecules 
\cite{Herrera2010b,Herrera2011,Lesanovsky2012,Herrera2013}, mobile impurities in Bose and Fermi gases \cite{Lampo2017,Mehboudi2018,Scazza2022},
ultracold  dipolar and Rydberg atoms
\cite{Wuster2011,Hague2012,Hague2014,PhysRevLett.114.113002,Plodzien2018,Camargo2018,PhysRevA.107.032808, Magoni2022,DiLiberto2022topologicalphonons}, to quantum dots on a carbon nanotube \cite{Pistolesi2021}.
Although each of these platforms possesses its unique strengths and benefits, recently there has been an exceptional outburst of interest in quantum simulation and computation with Rydberg atoms, which provide a remarkable level of flexibility for executing quantum operations and constructing quantum many-body Hamiltonians \cite{Morgado2021}. While the latter can contribute to our comprehension of the static properties of many-body systems, their main benefits are centered around exploring the complex dynamics displayed by these systems. In particular, in the context of polarons, it has been demonstrated that the dipole-dipole interactions between distinct Rydberg-dressed states can result in coherent quantum transport of electronic-like excitations \cite{Wuster2011}, which can further be coupled to optical phonons \cite{Hague2012}. The paradigmatic one-dimensional topological   Su-Schrieffer-Heeger (SSH) model \cite{Su1979} describing the soliton formation in long-chain polyacetylene due to excitation-phonon coupling, has been realized in Rydberg arrays \cite{doi:10.1126/science.aav9105, Weber_2018, Lienhard:19}.

In this paper, we continue along this path and present theoretical studies of an implementation of a microscopic model featuring the interplay of Su-Schrieffer-Heeger (SSH) and Fröhlich electron-phonon coupling mechanisms between optical phonons and excitations, under the influence of an external force and disorder.
In particular, we focus on the directional transport of an excitation interacting with phonons. We indicate an excitation-phonon coupling regime where the competition between Bloch oscillations and interactions results in the coherent transport of a well-localized wave packet over a long distance. We show the robustness of such a coherent transport of well-localized wave packets to the on-site random potential, indicating that a relatively strong disorder does not affect significantly the transport properties. Moreover, for completeness, we consider also excitation coupling to acoustic phonons.

The paper is divided into three parts. In the first part, Sec.~\ref{sec:Model}, we describe the physical setup and derive the effective Hamiltonian in Rydberg-dressed atomic arrays. The second part, described in Section~\ref{sec:experiment}, focuses on the dynamics of the system under experimentally relevant parameters. In this section, we observe the macroscopic transport of the center of mass and a transition between Bloch oscillations and moving polaron regimes. In the third part, Sec.~\ref{sec:phase_diagram}, we comprehensively analyze the previously derived microscopic model, which exhibits a rich phase diagram due to the interplay of two different electron-phonon coupling mechanisms. Finally, we compare the behavior of excitations with acoustic and optical phonons and demonstrate the robustness of our results.

\section{The model and its Hamiltonian}
\label{sec:Model}

We consider a one-dimensional chain of $N$ equidistant Rydberg atoms with lattice constant $x_0$ and positions $x_j = j x_0$, confined in a periodic trap, implemented either by an optical lattice  \cite{Anderson2011,Macr2014}, an optical tweezer array \cite{Browaeys2020,Kaufman2021,Wilson2022}, a Rydberg microtrap \cite{Leung2014}, or a
painted potential \cite{Henderson2009}. We assume that the spatial motion of the atoms is suppressed by the strong confinement of each Rydberg atom in local potential minima. Although the atomic motion is frozen, it is remarkable that such a Rydberg system can display highly non-trivial dynamics. 
In particular, the induced dipole-dipole interactions between distinct Rydberg-dressed states can lead to the emergence of coherent quantum transport of electronic-like excitations \cite{Wuster2011}.   
In the following, we first briefly repeat the derivation of the Hamiltonian that characterizes the dynamics of single excitations \cite{Wuster2011}. The purpose of this recap is to modify the setup in order to incorporate nearly arbitrary on-site potential terms.
Next, after introducing phonons into the system \cite{Hague2012}, 
we derive an effective nearest-neighbor Hamiltonian  that includes two  excitation-phonon coupling terms, which we comprehensively study in the forthcoming sections, focusing on the dynamics in the presence of an external constant field.

\subsection{Single excitation Hamiltonian in arbitrary potentials}
\label{sec:setup_a}
We assume that each Rydberg atom  can be initially found in one \ak{of} the ground state hyperfine levels, $|g\rangle$ or $|g'\rangle$. By applying far-detuned dressing laser fields, with effective Rabi frequencies $\Omega_{s}$, $\Omega_{p}$ and detunings $\Delta_{s}$, $\Delta_{p}$ respectively, these two hyperfine states can be coherently coupled to selected highly excited Rydberg states, $|s\rangle$ or $|p\rangle$, with principal quantum number $n\gg 1$ and different angular momenta. Consequently, each atom can be found in one of the two Rydberg dressed states \cite{Ates2008, Wuster2011,Pohl2014,Buchler2010,Rost2014,Zeiher2016}, which are a slight admixture of Rydberg states to the atomic ground states,
\be
|0\rangle_j \approx |g\rangle_j + \alpha_s |s\rangle_j \quad \mbox{or} \quad 
|1\rangle_j \approx |g'\rangle_j + \alpha_p |p\rangle_j ,
\ee
with $\alpha_{s/p} = \Omega_{s/p}/[2\Delta_{s/p}]$ and $j$ denoting the position of an atom.  Treating $\alpha_{s}$, $\alpha_{p}$ as perturbation parameters in van Vleck perturbation theory, W\"uster at al. \cite{Wuster2011} have shown that the dipole-dipole interaction can exchange the internal states of a neighboring pair, e.g. $|1\rangle_1  |0\rangle_2 \rightarrow |0\rangle_1  |1\rangle_2 $. This process can be viewed as a hopping of an excitation from $j=1$ to $j=2$ lattice site, which conserves the number of excitations. 

The perturbation analysis can be extended to a chain of $N$ atoms, where the effective Hamiltonian in the single excitation manifold (up to the fourth order in $\alpha_{s}$ and $\alpha_{p}$) reads \cite{Wuster2011,Hague2012}
\be
\label{Hamiltonian_alltoall}
\hat H_0 =   \sum_j \hat n_j (E_2 + E_4+  A_j) +  \sum_{j,k}  A _{jk} \hat a_j^\dagger \hat a_k,
\ee
where $\hat a_j$  ($\hat a^\dagger_j$) denote an annihilation (creation) operator of excitation on site $j$,  while 
\begin{subequations}
\label{eq:coefficents}
\begin{align} 
\label{Aj}
  A _j &= \hbar \alpha_s^2 \alpha_p^2 \left(\sum_{k \ne j} \frac{1}{1-\bar U_{kj}^2} \right) (\Delta_s + \Delta_p), \\ 
\label{Ajk}
  A_{jk} &= \hbar \alpha_s^2 \alpha_p^2  \frac{\bar U_{jk}}{1-\bar U_{jk}^2} (\Delta_s + \Delta_p),
\end{align}
\end{subequations} 
with $\bar U_{jk} = C_3/[\hbar \left| x_i- x_j\right|^{3} (\Delta_s + \Delta_p)]$ and $C_3$ quantifying the transition dipole moment between the
Rydberg states,
describe perturbative dipole-dipole interactions.
Finally, $E_2$ and $E_4$ are constant energy shifts of the second and fourth order, respectively,
\begin{subequations}
\label{eq:E1E2}
\begin{align} 
\label{E2}
E_2 /\hbar &= (N-1)\alpha_s^2 \Delta_s + \alpha_p^2 \Delta_p,\\ 
\label{E4}
E_4 /\hbar&= (N-1) \alpha_s^4  \Delta_s+ \alpha_p^4 \Delta_p \nonumber  \\ 
&   +  (N-1)\alpha_s^2 \alpha_p^2 (\Delta_s+ \Delta_p)  . 
\end{align}
\end{subequations} 
Although in principle constant energy terms could be always ignored as they do not contribute to the dynamics of excitations, let us consider now a scenario where the Rabi frequency $\Omega_p$ depends on the atomic position on the lattice, i.e., we assume that  
\be
\Omega_p \rightarrow  \Omega_p(j) \equiv \Omega_p \left[1 + \delta\Omega(j) \right],
\ee
where $\delta\Omega(j)$ is arbitrary, but small correction of the order $(\alpha_{p/s})^2$. With this assumption, and by retaining terms up to the fourth order, the effective Hamiltonian in  Eq.~\eqref{Hamiltonian_alltoall} acquires an additional term, namely
\be\label{ham_with_force}
\hat H =  \hat H_0 + \hbar \alpha_p^2  \Delta_p \sum_j \delta \Omega(j) \hat n_j \;. 
\ee
Because the term proportional to $\alpha_p^2\delta\Omega(j)$ is of the same order as $A_j$, it can be incorporated into the definition of $A_j$ in Eq.~\eqref{Aj}.
With this simple modification, we have gained a position-dependent effective potential term that can strongly affect the dynamics of excitations. Although the potential term can be tailored almost arbitrarily, from now on  we consider  one of its simplest forms, i.e., we choose
\be
\delta\Omega(j)   = 2 \alpha_s^2 \left(F j + \epsilon_j  \right).
 \ee
 The first term in the parentheses, being linearly proportional to position $j$,  emulates the presence of a constant external field $F$. 
 The second term, with  $\epsilon_j$ being a random variable, gives rise to the on-site potential disorder. 
Note that both terms lead to localization of the excitation either due to Stark localization \cite{Glueck02}
in a constant tilt, $F$, or Anderson localization \cite{Evers08} due to random $\epsilon_j$. As explained in the next part, the situation is not so straightforward. 
 
\subsection{Excitation-phonon Hamiltonian}
\label{sec:setup_b}
In this part, we  relax our previous assumption that the atoms of the array are completely immobile.  Although we still assume that no atom can move through the lattice, we now let them vibrate in the vicinity of their local equilibrium points. This will affect, as we shall see, the dynamics of excitations. We consider now a scenario where
an atom
in the $j$-th lattice site and with mass $m$ may oscillate with a frequency $\omega_0 = \sqrt{k/m}$ inside a local potential well, that can be approximated by a quadratic potential
\be
\frac{k}{2} (x-j x_0)^2 \equiv \frac{k x_0^2}{2} (u_j)^2,
\ee 
with $k$ being the force constant and where  $u_j$ denotes dimensionless distortion from the local equilibrium position.
The motion of atoms can be quantized $u_j \rightarrow \hat u_j$ and  described by a simple quantum harmonic oscillator. This vibrational motion is responsible for the distortion of an atomic array and can be considered as a phonon. Since the Hamiltonian of the previous section describing the motion of single excitations strongly depends on the position of atoms,  phonons can propagate through space due to the coupling to excitations. Before proceeding to derive the effective Hamiltonian of the system with phonon-excitation coupling, for clarity and simplicity we assume that
\be
\alpha \equiv \alpha_s=\alpha_p , \quad \Delta \equiv \Delta_s=\Delta_p .
\ee
Moreover, from now on we also fix the time and energy scales and go to the dimensionless units by dividing all the energy scales by  $2\hbar \alpha^4 \Delta$.

Although the setup described in Section~\ref{sec:setup_a}  admits only dispersionless optical phonons that correspond to local vibrations of atoms around local minima,  we consider here two different types of phonons. We proceed by writing the phononic Hamiltonian explicitly in terms of  the dimensionless position and momentum operators $\hat{u}_j$,  $\hat{p}_j$  of local distortions,
\be\label{ham_phonons}
 \hat{H}_\text{ph}  = 
 \sum_j \frac{\hat{p}^2_j}{2 m_{\text{eff}}} + \frac{m_{\text{eff}}\omega_{\text{eff}}^2}{2} (\hat{u}_j - \eta \hat{u}_{j-1})^2,
\ee
with the effective dimensionless mass,
\be\label{eq:eff_mass}
m_{\text{eff}} = 2 m  x_0^2 \alpha^4 \Delta / \hbar,
\ee
and the effective oscillator frequency,
\be\label{eq:eff_omega}
\omega_{\text{eff}} =\omega_0 / (2 \alpha^4 \Delta), \quad \omega_0 = \sqrt{k/m},
\ee
where $\omega_0$ is the bare frequency. By changing the parameter $\eta$ in Eq.~\eqref{ham_phonons}, diverse phonon types can be achieved. In particular, 
$\eta=0$ corresponds to the aforementioned local vibrations (i.e., dispersionless optical phonons), and $\eta=1$ describes acoustic phonons.
These two phonon types are characterized by the dispersion relation
\begin{equation}\label{eq:disp_relation}
\epsilon_{q} = 
\begin{cases}
               \omega_{\text{eff}} ,      & \mbox{(optical phonons, $\eta = 0$)} \\
                2 \omega_{\text{eff}} \left|\sin(q x_0/2)\right|,   & \mbox{(acoustic phonons, $\eta = 1$)}
            \end{cases},
\end{equation}
which can be readily found by writing the phononic Hamiltonian \eqref{ham_phonons} in terms of its eigenmodes,
\be
 \hat{H}_\text{ph}  = \sum_q \epsilon_{q} \left(\hat{b}^\dagger_q\hat{b}_q+\frac{1}{2}\right),
\ee
where $\hat{b}^\dagger_q$ ($\hat{b}_q$) creates (annihilates) the phonon with quasi-momentum $q$, and are related to the  local dimensionless momentum and position operators $\hat{p}_i$, $\hat{u}_i$ of distortion  by 
\begin{equation}
\begin{split}
 \hat{u}_j &= \sum_q \frac{1}{\sqrt{2 N  \epsilon_{q} m_{\text{eff}}}}(\hat{b}_q+ \hat{b}^\dagger_{-q})e^{iqjx_0},\\
 \hat{p}_j &= -i\sum_q \sqrt{\frac{  \epsilon_{q} m_{\text{eff}}}{  2 N }}(\hat{b}_q - \hat{b}^\dagger_{-q}) e^{iqjx_0}.
\end{split}
\end{equation}

\begin{figure}
    \centering
\includegraphics[width=1\linewidth]{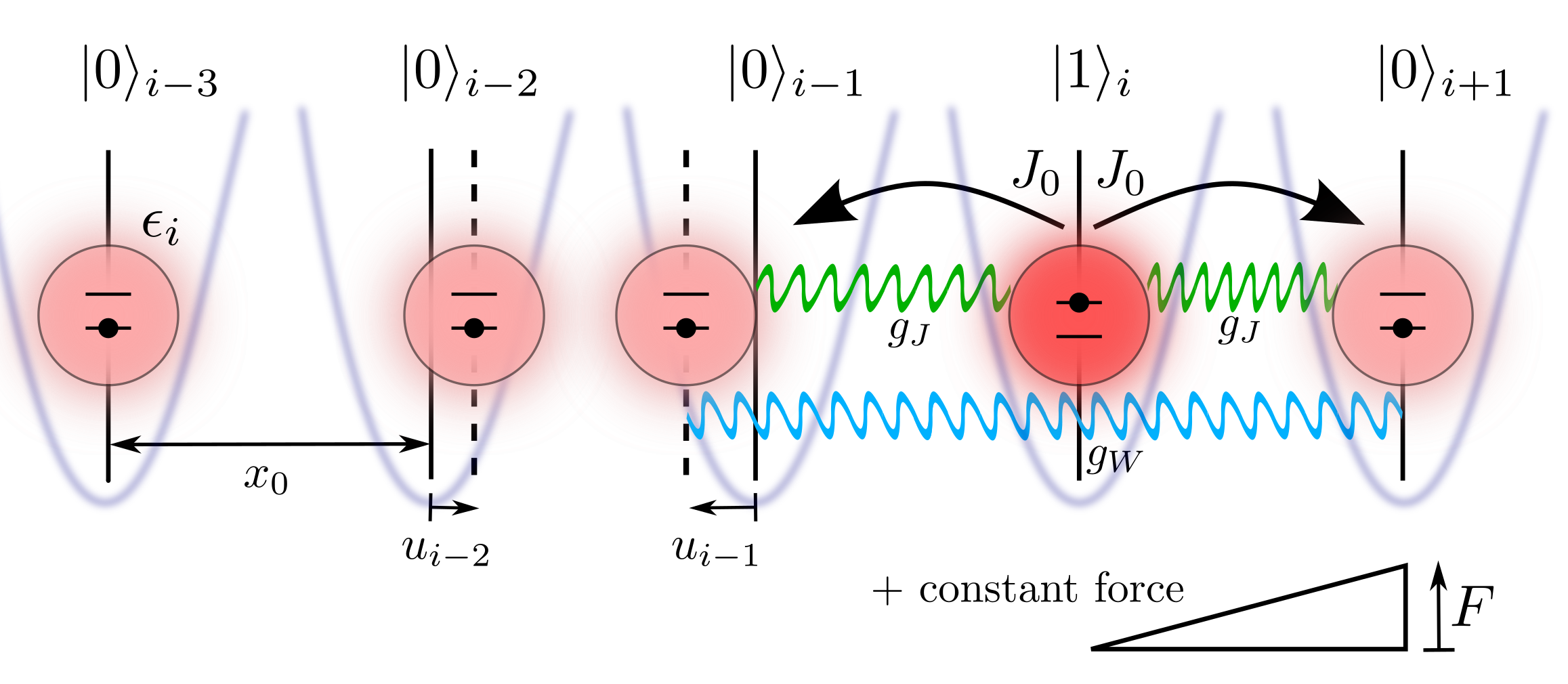}
    \caption{Schematic illustration of all processes in the effective Hamiltonian $\hat{H}_\text{eff}$ in Eq.~\eqref{eq:Hamiltonian_final}, describing the dynamics of a single excitation in a one-dimensional array of Rydberg atoms located at  $x_0(j + u_j)$, with $u_j$ being  a dimensionless distortion from an equilibrium position. $\hat{H}_\text{ex}$ describes a bare hopping (in the limit $u_j\rightarrow0$) of an excitation at site  $j$  to its neighboring sites with amplitude $J_0$ in the presence of a constant force $F$ and on-site disorder $\epsilon_j$, see Eq.~\eqref{H_ex}.  
     The effective hopping and on-site potential is further modified by the phonon couplings $g_J$ and $g_W$, respectively, see Eq.~\eqref{eq:phonon-coupling}. 
    }
    \label{fig:fig1}
\end{figure}

Having discussed the phononic degrees of freedom, we can now write the fully effective Hamiltonian governing the motion of single excitations coupled to phonons. The derivation is straightforward and requires: \textit{(i)} the expansion of the position-dependent coefficients [given by Eq.~\eqref{eq:coefficents}] in the Hamiltonian \eqref{ham_with_force} of the previous section up to the first order in $ \hat u_j $, and \textit{(ii)} dropping the next-to-nearest neighbor contributions \cite{foot_kappa}.
By following these steps, we obtain the effective excitation-phonon Hamiltonian~[cf.~Fig.~\ref{fig:fig1}], which consists of four parts, i.e.,
\begin{equation}\label{eq:Hamiltonian_final}
 \hat{H}_\text{eff}  = \hat{H}_\text{ph}    + \hat{H}_\text{ex} + \hat{H}_\text{J} + \hat{H}_\text{W},
\end{equation}
where $\hat{H}_\text{ph}$ is the phononic Hamiltonian, Eq.~\eqref{ham_phonons}, 
\be \label{H_ex}
 \hat{H}_\text{ex}  =   J_0(\hat{a}^\dagger_{j+1}\hat{a}_j + \mbox{H.c.}) + \sum_j (j F  +\epsilon_j)\hat{a}^\dagger_j\hat{a}_j , 
\ee
describes excitations with the hopping amplitude 
\be
\label{J0}
J_0 = \kappa/(1-\kappa^2), \quad \kappa = C_3/(2\hbar \Delta x_0^3),
\ee
experiencing an external  constant force $F$, and a local on-site disorder $\epsilon_j$.  Finally,
\begin{subequations}
\label{eq:phonon-coupling}
\begin{align} 
\label{SSH}
   \hat{H}_\text{J}  &= g_J\sum_j (\hat{u}_{j+1} - \hat{u}_{j}) \hat{a}^\dagger_{j+1}\hat{a}_j +  \mbox{H.c.}  \, , \\
\label{Frohling}
   \hat{H}_\text{W}  &= g_W \sum_j (\hat{u}_{j+1} - \hat{u}_{j-1})\hat{a}^\dagger_j\hat{a}_j,
\end{align}
\end{subequations} 
are the notable SSH and Fr\"ohling Hamiltonians \cite{Su1979,Franchini2021}, respectively, that correspond to two different mechanisms of excitation-phonon couplings, 
with dimensionless coupling parameters
\begin{subequations}
\label{gjgw}
\begin{align} 
\label{gj}
g_J &= -3\kappa(1+\kappa^2)/(\kappa^2-1)^2, \\
\label{gw}
g_W &= -6\kappa^2/(\kappa^2-1)^2 .
\end{align}
\end{subequations}

\subsection{Equations of motion}
\label{sec:setup_c}

The full numerical analysis of the polaron dynamics on the many-body level is one of the most challenging computational tasks, due to the non-conserved total number of phonons in the system, which prevents it from working in a restricted, fixed particle-number Hilbert space sector of the phononic degrees of freedom. Additionally, even without a force $F$ the effective Hamiltonian of the systems \eqref{eq:Hamiltonian_final} depends, in principle, on many parameters, namely $J_0, g_W, g_J, \omega_{\rm eff}$ and $m_{\rm eff}$, making the full analysis of the system even more challenging. 

To analyze the dynamical properties of the considered system, in the following we make the semiclassical approximation by applying the Davydov Ansatz \cite{Davydov1969,Zhao1997,Zhou2015,Zhou2016,Huang2017a,Huang2017b,Zhao2021}, which relies on two fundamental assumptions: (i) treating phononic oscillations classically and (ii) representing solutions as separable states without entanglement between quantum-like excitations and classical-like phonons. In other words, we assume that phonons are in a coherent state and that the full wave function is a product state of the excitation and coherent phonons part, as
\begin{equation}\label{Davydov_anzatz}
 | \Psi (t)\rangle = \left(\sum_j\psi_j(t)\hat{a}_j^\dagger\right)\otimes\left( e^{-i \sum_j\left[  u_j(t)\hat{p}_j-p_j(t)\hat{u}_j\right]}\right)|\mathtt{vac}\rangle,
\end{equation} 
where $|\psi_j(t)|^2$ is a probability of finding an excitation at a site $j$, $u_j(t)$ and $p_j(t)$ are expectation values of phononic position and momentum operators. The equation of motion for $\psi_j(t)$ and $u_j(t)$ can be subsequently derived from a classical conjugate variable Heisenberg equations of motions using the generalized Ehrenfest theorem, see, for example, Ref.~\cite{Georgiev2019}.
By following these steps, we obtain a closed set of coupled differential equations for the excitation amplitude $\psi_j(t)$ and classical field $u_j(t)$. The equations can be written in a concise form, as
\begin{subequations}
\label{eq:DavidovEqs}
\begin{align} 
\label{eq:DavidovEq1}
i \dot{\psi_j} &=  J_j \psi_{j+1} + J_{j-1} \psi_{j-1} + W_j \psi_j , \\
\label{eq:DavidovEq2}
    \ddot{u}_j  &= -\omega_{\text{eff}}^2 \, {\cal D}[\{u_j\}] + {\cal S}[\{\psi_j\}],
\end{align}
\end{subequations}
where the effective potential experienced by an excitation  $W_j(t)$,  and the effective hopping amplitude $J_j(t)$ are both time-dependent functions due to the coupling to the gradient of the phononic field $u_j(t)$, i.e.
\begin{equation}
    \begin{split}
        W_j(t) &= j F +\epsilon_j + g_W[u_{j+1}(t) - u_{j-1}(t)]\\
        J_j(t) &= J_0 + g_J[u_{j+1}(t) - u_j(t)].
    \end{split}
\end{equation}
As such, both $W_j(t)$ and $J_j(t)$ are responsible for the self-trapping of an excitation.  Similarly, the phononic equation \eqref{eq:DavidovEq2} also depends on the excitation amplitude  $\psi_i(t)$ through the ${\cal S}[\{\psi_j\}]$ operator, given by
\be
\begin{split}
{\cal S}[\{\psi_j\}]   =  -  \frac{g_W}{m_{\text{eff}}} (|\psi_{j+1}|^2 - |\psi_{j-1}|^2)\\ 
		   - \frac{g_J}{m_{\text{eff}}}[\psi_j^*(\psi_{j+1} - \psi_{j-1}) +  \mbox{c.c.}],
\end{split}
\ee
which acts as a time-dependent source for the phonon field  $u_j(t)$.
Finally, the phononic dispersion relation, given by Eq.~\eqref{eq:disp_relation}, is necessarily present in the phononic equation through the ${\cal D}[\{u_j\}]$ operator,
\begin{equation}
{\cal D}[\{u_j\}]= 
\begin{cases}
               u_j ,      & \eta = 0,  \\
              2u_j - u_{j+1}-u_{j-1},  &  \eta = 1, 
            \end{cases}\;,
\end{equation}
which introduces a crucial difference in the propagation of optical ($\eta=0$) and acoustic ($\eta=1$) phonons \cite{Kittel2004}, which we investigate in the next sections.

\subsection{Analysed observables}
\label{sec:setup_d}

Throughout this article we choose the initial conditions $\psi_j(0) = \delta_{j,0}$ and  $u_j(0) = \dot u_j(0) = 0 $ for the equations of motion, Eq.~\eqref{eq:DavidovEqs},  that correspond to a single excitation on a central lattice site and initially unperturbed lattice. Without a phonon-coupling and for $F=0$, these initial conditions simply correspond to a quantum particle that spreads symmetrically in both lattice directions characterized by a constant Lieb-Robinson velocity \cite{Lieb1972}, so that its center of mass remains localized at the initial position. Contrary to the classical case, a quantum particle on a lattice will not even move in the presence of a constant force~$F$, but  instead,  it starts to perform Bloch oscillations \cite{Bloch1929}.  The situation is different in interacting systems, either in a case of particle-particle interactions \cite{Wiater2017}, which may further lead to disorder-free many-body localization \cite{vanNieuwenburg2019,Schulz19,Taylor20,Guo20,Yao20b,Chanda20c,Yao21,Scherg21,Morong21,Kohlert23}, or in the presence of phonons, which 
can induce transient polarons at the end of Bloch oscillation periods \cite{Huang2017a,Nazareno2016} (see also Ref.~\cite{Li2006}). 

In this study, we investigate how the propagation of a single excitation is influenced by the two competing phonon-coupling mechanisms under the applied, constant force. Specifically, we aim at answering the two following questions: \textit{(i)}~how much does the excitation spread due to the coupling with phonons, and \textit{(ii)}~does its center of mass move in the presence of the constant force  $F$? In order to respond to these questions we focus on three simple observables that can be calculated based on the local density measurements. First, we consider the participation ratio (PR), defined as \cite{Bell1970}:
\be\label{eq:PR}
\mbox{PR}(t) = \bigg(\sum_j\left|\psi_j(t)\right|^4\bigg)^{-1},
\ee
where we have assumed a unit normalization of the wavefunction $\sum_j|\psi_j|^2=1$.
The participation ratio PR is equal to $1$ where excitation is localized on a single lattice site and equals $N$ when is completely delocalized over the whole lattice.
The second observable is the center of mass position of the wave packet, i.e.,
\be\label{eq:xcm}
x(t) = \sum_{j=-N/2}^{N/2} j \left|\psi_j(t)\right|^2.
\ee
Moreover, in some cases, analyzing the ratio of the two quantities mentioned above can provide valuable insights. We define this ratio, denoted as $\xi$, as:
\be\label{eq:xi}
\xi(t) = \frac{|x(t)|}{\mbox{PR}(t)}.
\ee
$\xi $ is a quantity ranging from 0 to $\xi_{\textrm{max}} = N/2$. The maximum value $\xi_{\textrm{max}}$ corresponds to a moving, maximally-localized, non-dispersive solution that has reached the boundary of the system. As such,  $\xi$ can be viewed as an indicative measure for selecting well-localized solutions moving in one direction. 

Finally, it is worth  mentioning that it is often not necessary to analyze the entire time range of the above observables. In fact, to discern various dynamic behaviors, it is usually sufficient to look at $\mbox{PR}(t)$, $x(t)$ and $\xi(t)$ at the final evolution time $t_f \gg 1$. For example, large $\mbox{PR}(t_f)$ (relative to the system size $N$) suggests that excitation is not stable and has delocalized over a lattice. 

\section{Polaron dynamics: experimental considerations}
\label{sec:experiment}

In this section we elaborate on the results of the previous sections and study the dynamics of a Rydberg excitation under the presence of the external force $F$, solving the equations of motion for a physically relevant range of parameters.  
The effective Hamiltonian \eqref{eq:Hamiltonian_final} of the system relies on several effective, dimensional parameters, including $m_{\text{eff}} = 2 m  x_0^2 \alpha^4 \Delta / \hbar$, $\omega_{\text{eff}} =\omega_0 / (2 \alpha^4 \Delta)$, as well as $J_0$, $g_J$, $g_W$, given by Eq.~\eqref{J0} and Eqs.~\eqref{gjgw}.  However, it is worth noting that the latter three parameters are not independent within our setup, and their values are determined by a single parameter $\kappa = C_3 / (2\hbar \Delta x_0^3)$. This provides us with significant flexibility in selecting appropriate physical parameters for our convenience.  We stress, that the proposed quantum simulator allows for simulating excitation coupled to optical phonons only.

In the following, we choose the highly excited Rydberg states $|s\ra $, $|p\ra$  of Rubidium-87 with principal quantum 
number $n=50$ and angular momentum equal to 0 or $\hbar$, for which  $C_3 = 3.224$ GHz$\times\mu m^{-3}$. We fix the lattice spacing $x_0 = 2\, \mu \mbox{m}$, and the local trap frequency $\omega_0 =20$~kHz.  In the numerical simulations, we vary the dimensionless parameter $\kappa$ between $0.80-0.86$, which is equivalent to the change of the detuning $\Delta \sim 234-252$ MHz, and corresponds to the dressing parameter $\alpha \sim 0.04$. Importantly, by increasing $\kappa$ we also increase the phonon coupling strength from around $g_J/m_{\textrm{eff}} \sim g_W/m_{\textrm{eff}}  \sim-4.5   $ to $g_J/m_{\textrm{eff}} \sim g_W/m_{\textrm{eff}}  \sim-8   $.  Furthermore, we remind the reader that in our setup only the optical phonons (i.e., dispersionless vibrations) are experimentally relevant and, therefore, in this section we set $\eta=0$.  Finally, we fix the value of the force at $F=0.2$, and we choose  the  system size to $N=401$.

\begin{figure}[t!]
    \centering
    \includegraphics[width=\linewidth]{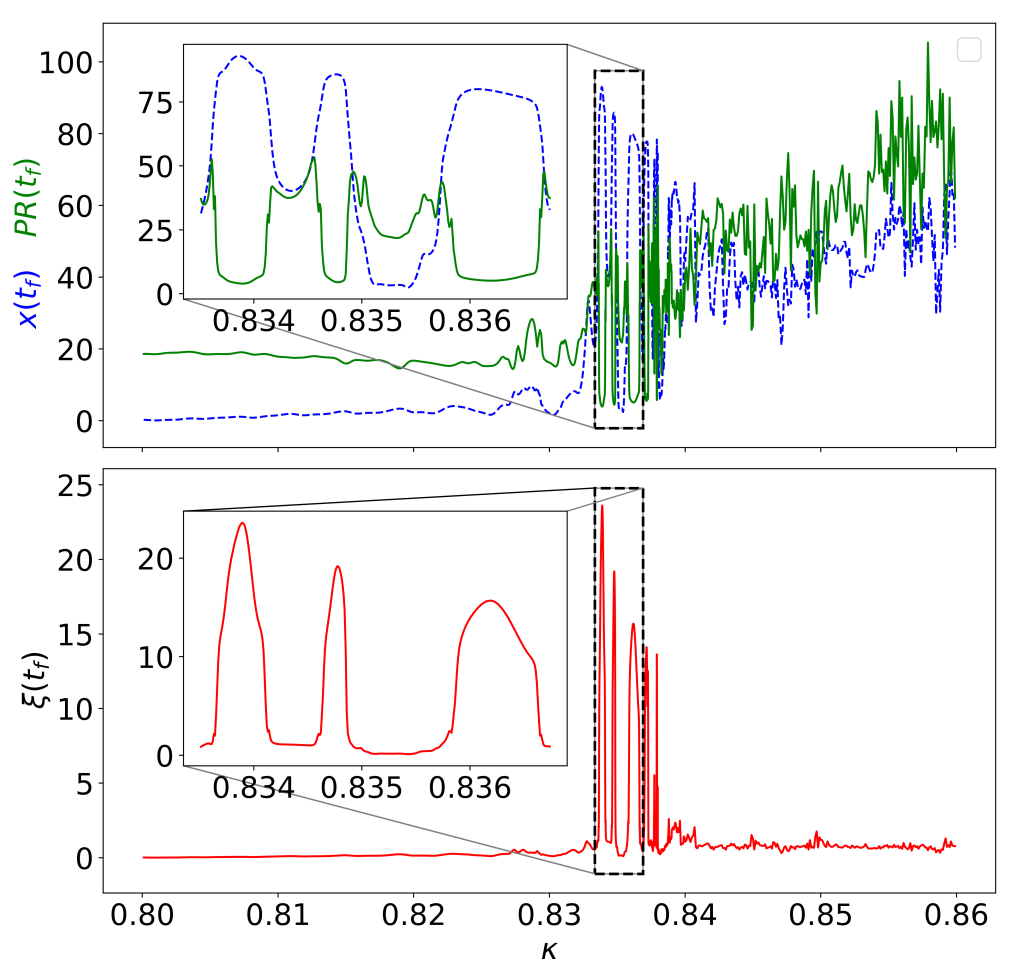}
\caption{
The top panel illustrates the center of mass motion $x(t)$ of an excitation dashed blue line) under a constant, external force $F=0.2$, and the corresponding participation ratio $\text{PR}(t)$ (solid green line), evaluated at the final evolution time $t_{f}$, and  plotted as functions of the dimensionless parameter $\kappa$
(see Eq.~\eqref{eq:Hamiltonian_final} and definitions below it). 
In the bottom panel, the ratio $\xi = |x|/\text{PR}$ is shown. The peaks in the plot correspond to parameter regimes where a well-localized excitation is transferred under the influence of a constant force $F$. All physical parameters have been chosen with careful consideration of their experimental relevance, as discussed in the main text. The time evolution range is $t \in [0, t_f]$, where the final time $t_{f}$ is chosen as $t_{f} = 2.1 \, T_B = 4.2\, \pi/F \approx 66$. 
}
    \label{fig:fig2}
\end{figure}

In order to characterize the transport properties of an excitation $\psi_i(t)$, in the top panel of Fig.~\ref{fig:fig2} we plot its center of mass position $x(t)$ and the corresponding  participation ratio $\text{PR}(t)$, see Eqs.~\eqref{eq:PR}-\eqref{eq:xcm} for the respective definitions. In the bottom panel, we additionally illustrate the ratio  $\xi = |x|/\text{PR}$.
All these quantities are plotted as a function of $\kappa$, at a fixed time $t_f=2.1 \, T_B \approx 66$, where $T_B = 2\pi/F $ is the Bloch oscillation period. We find that up to $\kappa \sim 0.83$ both $x(t_f)$ and $\text{PR}(t_f)$ are small (relative to the system size $N$) which corresponds to the Bloch oscillation-like dynamics where the phonon-influence is minimal. In contrast, phonons play important role above $\kappa \sim 0.83$ where the system dynamics is quite sensitive to the choice of microscopic parameters. 
Within the chaotic-like regime, the typical Bloch oscillation dynamics is completely disrupted, as the majority of solutions become delocalized across the lattice, leading to large values of $\text{PR}(t)$. However, amidst this chaotic behavior, we also discover intervals of stability, characterized by peaks of $\xi(t_f)$, where a substantial portion of the wave packet becomes well-localized and exhibits near-constant velocity motion. 

\begin{figure}[t!]
    \centering
    \includegraphics[width=1\linewidth]{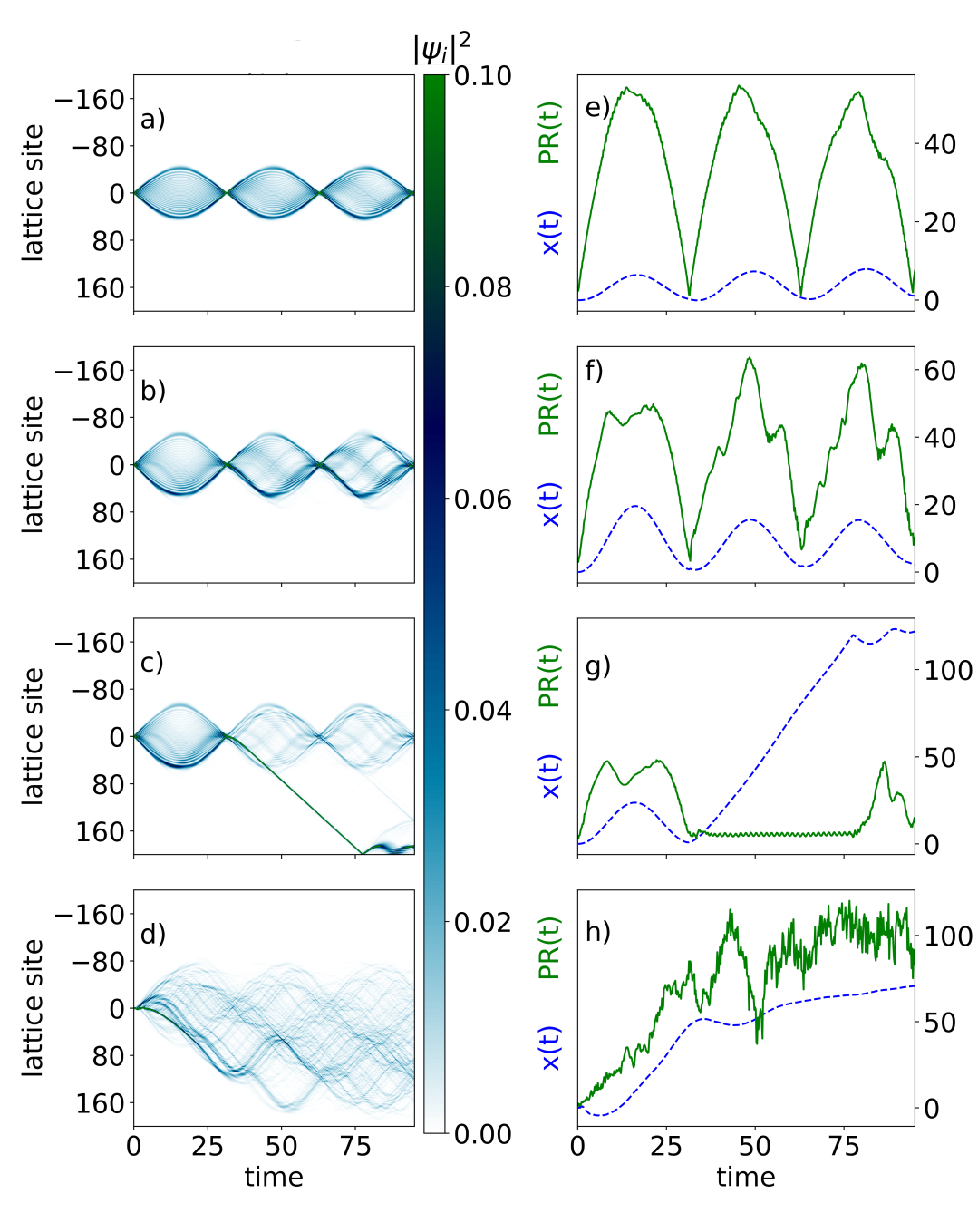}
    \caption{The panels illustrate the diverse dynamic behaviors observed in our study. The first column, panels (a)-(d), showcases the time evolution of the excitation density $|\psi_j(t)|^2$ (color encoded), while the second column, panels (e)-(h), illustrates the corresponding temporal changes in the center of mass position $x(t)$  (dashed blue lines) and the participation ratio PR$(t)$ (solid green lines).
In the first row ($\kappa = 0.8$), near-perfect Bloch oscillations are observed. However, upon closer examination, a subtle yet discernible asymmetry becomes apparent, as indicated by the non-zero value of $x(t)$. This asymmetry becomes more pronounced in the second row for a higher $\kappa$ value of 0.83.
The subsequent rows of Fig.~\ref{fig:fig3} provide insights into the time evolution of the excitation density for the specific cases of a well-localized wave-function ($\kappa=0.834$) and a spreading wave-function ($\kappa=0.86$). These distinct parameter regimes highlight the contrasting behavior and spatial characteristics of the excitations. All physical parameters are the same as in Fig.~\ref{fig:fig2}.}
    \label{fig:fig3}
\end{figure}

We illustrate those different dynamical behaviours in Fig.~\ref{fig:fig3}, where the first column, i.e., panels (a)-(d), show the time evolution of the excitation density $|\psi_j(t)|^2$, while the second column [panels (e)-(h)] illustrates the corresponding time evolution of the center of mass position $x(t)$ and the participation ratio PR$(t)$.  In the first row ($\kappa = 0.8$), we observe almost perfect Bloch oscillations. However, upon closer examination, a subtle asymmetry becomes apparent, which is evident by a non-zero $x(t)$. The asymmetry is enhanced for a higher $\kappa = 0.83 $, as depicted in the second row of Fig.~\ref{fig:fig3}. Finally, the last two rows of Fig.~\ref{fig:fig3} illustrate the time evolution of the excitation density in the chaotic-like regime above $\kappa \sim 0.83$, cf. Fig.~\ref{fig:fig2}, where most of the solutions are delocalized over a lattice, as in  Fig.~\ref{fig:fig3}(d) for $\kappa = 0.86$. In contrast, in  Fig.~\ref{fig:fig3}(c) we illustrate a regular behaviour for $\kappa=0.834$, which lies inside one of the aforementioned stability windows. 
In this scenario, due to constructive interference after one Bloch oscillation period, a prominent portion of the wave function coalesces into a very narrow non-dispersive wave packet that moves with a nearly constant velocity.   Overall, Fig.~\ref{fig:fig3} offers a comprehensive visual representation of the dynamic phenomena investigated in this section, shedding light on the varying dynamical behaviors and properties of the system with increasing phonon interaction. 

\section{Dynamical phase diagrams of the effective Hamiltonian} 
\label{sec:phase_diagram}

\begin{figure}[t!]
    \centering
\includegraphics[width=1\linewidth]{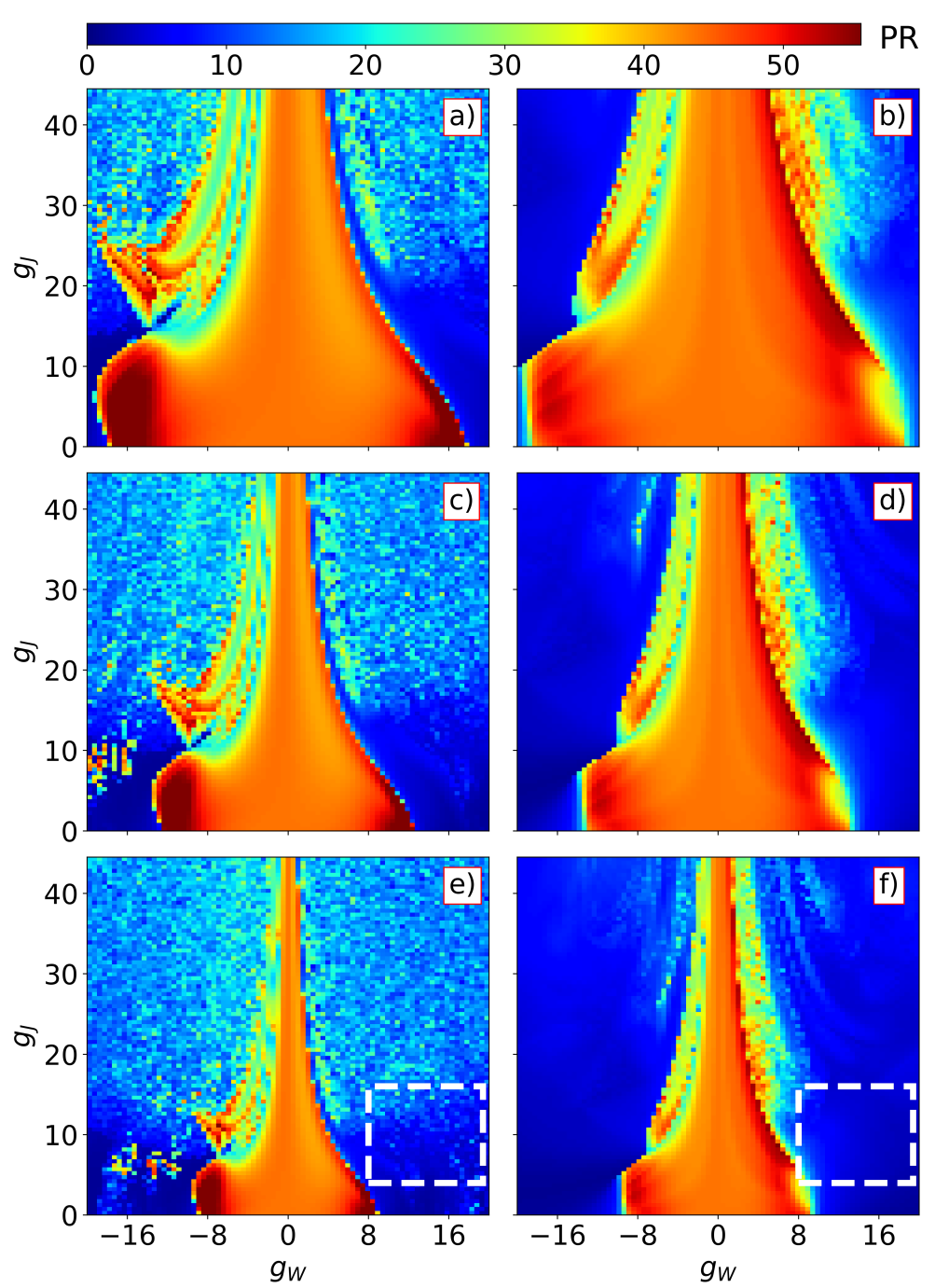}
    \caption{Color encoded participation ratio ${\rm PR}(t_f)$, Eq.~\eqref{eq:PR}, at the final evolution time for a broad range of coupling strengths, $g_J \in [0,45]$ and $g_W \in [-20,20]$, in the presence of optical (left column, $\eta = 0$) and acoustic phonons (right column, $\eta = 1$).
    Each row corresponds to a specific value of the effective mass $m_{\rm eff}$. Panels (a)-(b) correspond to $m_{\rm eff} = 2$, panels (c)-(d) correspond to $m_{\rm eff} = 1$, and
    panels (e)-(f) correspond to $m_{\rm eff} = 0.5$. Across all panels, we observe a mixture of extended states (warm colors) and well-localized, non-spreading wave solutions (dark blue colors). Moreover, decreasing effective mass $m_{\rm eff}$ narrows the delocalized phase. Despite some differences, both types of phonons exhibit qualitatively similar behavior, see the discussion in the main text. 
    The remaining parameters used for this analysis are $\omega_{\rm eff} = 10$, $J_0 = 1$.}
    \label{fig:fig4}
\end{figure}

In the previous sections, we have derived and then analysed a microscopic Hamiltonian \eqref{eq:Hamiltonian_final}, governing the dynamics of an excitation 
coupled to phonons through two different mechanisms, i.e., the SSH and Fr\"ohling Hamiltonians, see Eq.~\eqref{eq:phonon-coupling}. While maintaining a close connection to the experimental platform, it is important to note that in the considered Rydberg setup, the phonon coupling strengths $g_J$ and $g_W$ are not independent. Instead, they can both be expressed in terms of a single parameter $\kappa$, as demonstrated in Eq.~\eqref{gjgw}. Consequently, investigating the interplay between these two competing phonon-coupling mechanisms within the current Rydberg platform becomes challenging. To address this limitation and explore the complete phase diagram in a more general context, in this section, we treat $g_J$ and $g_W$ as completely independent and fix other parameters. In the initial phase, as described in Section \ref{sec:polaron_formation}, our primary objective is to identify a stable polaron regime. Specifically, we aim to find a regime in which an initially localized excitation does not spread during the course of time evolution. Subsequently, in Section \ref{sec:polaron_dynamics_disorder}, we demonstrate the existence of stable islands where polarons can exhibit non-dispersive motion when subjected to a constant force, even in the presence of substantial disorder.
Furthermore, in this part, we thoroughly examine the quantitative differences in dynamics of optical and acoustic phonons. 
In the following, we set the system size to  $N = 401$, and solve the equations of motions in a fixed time interval $t\in [0,t_f = 16.5] $.  Unless explicitly stated otherwise, we also set  $m_{\rm eff}=0.5$, $\omega_{\rm eff} = 10$, and $J_0 = 1$.

\begin{figure}[t!]
    \centering
\includegraphics[width=1\linewidth]{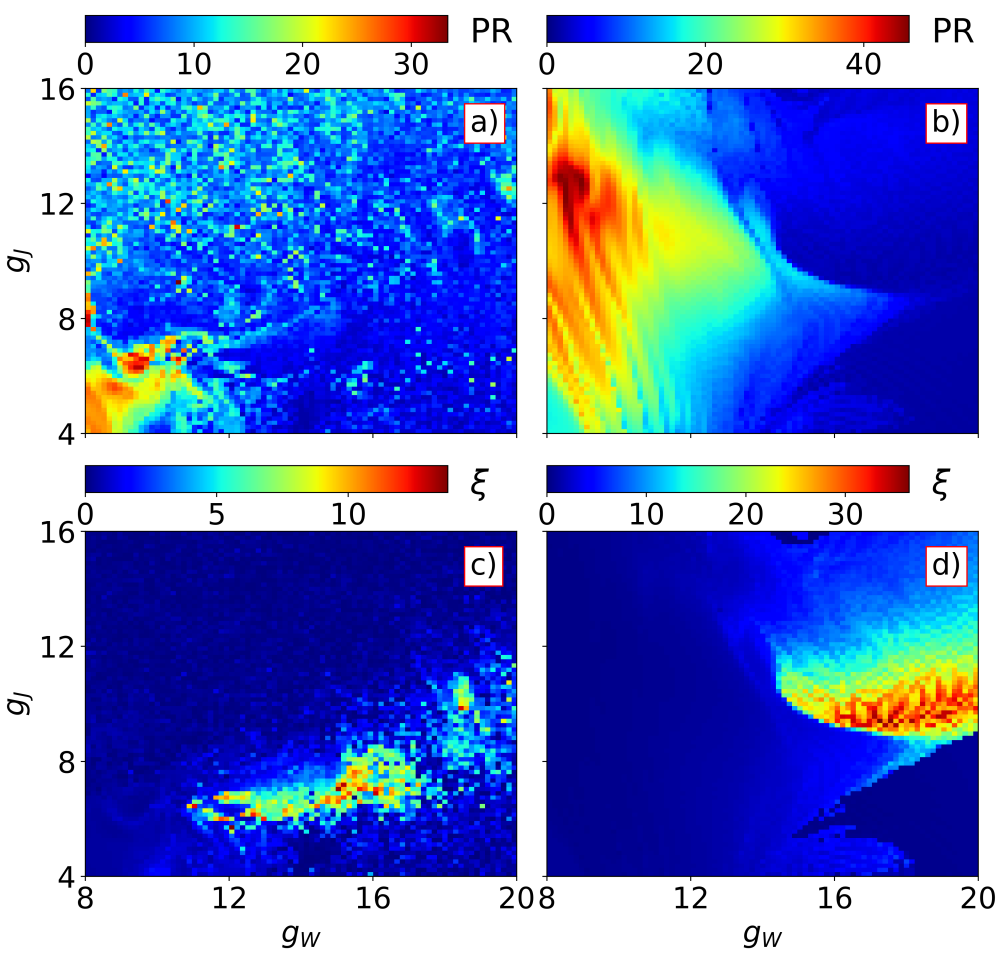}
    \caption{The influence of a constant force on the propagation of non-spreading solutions. Panels (a) and (b) depict the color encoded participation ratio, PR$(t_f)$, for optical and acoustic phonons, respectively, showing a shift in the boundary between extended and localized states due to the applied force. Panels (c) and (d) display color encoded $\xi(t_f)$, a measure for selecting well-localized solutions propagating in a single direction. Stable transport islands of such solutions are observed, indicated by warm colors. Panel (c) corresponds to optical phonons, while panel (d) corresponds to acoustic phonons. The remaining parameters used for this analysis are $F=0.2$, $m_{\rm eff}=0.5$, $\omega_{\rm eff} = 10$, $J_0 = 1$. While comparing with Fig.~\ref{fig:fig4} mind a shifted color scale.
    }
    \label{fig:fig5}
\end{figure}

\subsection{Polaron formation} 
\label{sec:polaron_formation}

In the preceding section, we have already witnessed the emergence of a non-dispersive, self-trapped polaron through the excitation-phonon coupling. Building upon this observation,  here we independently vary the two coupling strengths, $g_J$, and $g_W$, to identify a stable polaron regime. It is worth noting that the Hamiltonian of the system, as described by Eq.~\eqref{eq:Hamiltonian_final}, is invariant under the simultaneous transformation:  $u_j \rightarrow - u_j$, $g_J \rightarrow - g_J$, and $g_W \rightarrow - g_W$. Therefore, without loss of generality, we can assume $g_J\geq0$.

In Fig.~\ref{fig:fig4}, we present a phase diagram of the participation ratio PR calculated at the final evolution time for a broad range of values: $g_J\in [0,45]$ and $g_W\in [-16,20]$. Each panel of Fig.~\ref{fig:fig4} corresponds to distinct values of $m_{\rm eff}$ and $\eta$, as specified in the figure caption. 
In terms of the layout, the left (right) column corresponds to the optical (acoustic) phonons, and $m_{\rm eff}$ increases from top to bottom.
 In all panels of Fig.~\ref{fig:fig4}, we observe wide regions with both extended states (warm colors) and well-localized solutions (dark blue colors), with the latter corresponding to stable, stationary polarons.  We discover a non-trivial dependence of the participation ratio on both coupling strengths. 
Moreover, we find qualitatively similar behavior for both types of phonon, however, the acoustic phonons exhibit greater dynamic stability. This is evident from the presence of a chaotic-like region (the light blue dotted area), [compare with Fig.~\ref{fig:fig2} and see the discussion in Sec.~\ref{sec:experiment}]. Finally, we indicate that a decrease of effective mass $m_{\rm eff}$ stabilizes the excitation supporting localized polaron formation.

\begin{figure}[t!]
    \centering
\includegraphics[scale=0.33]{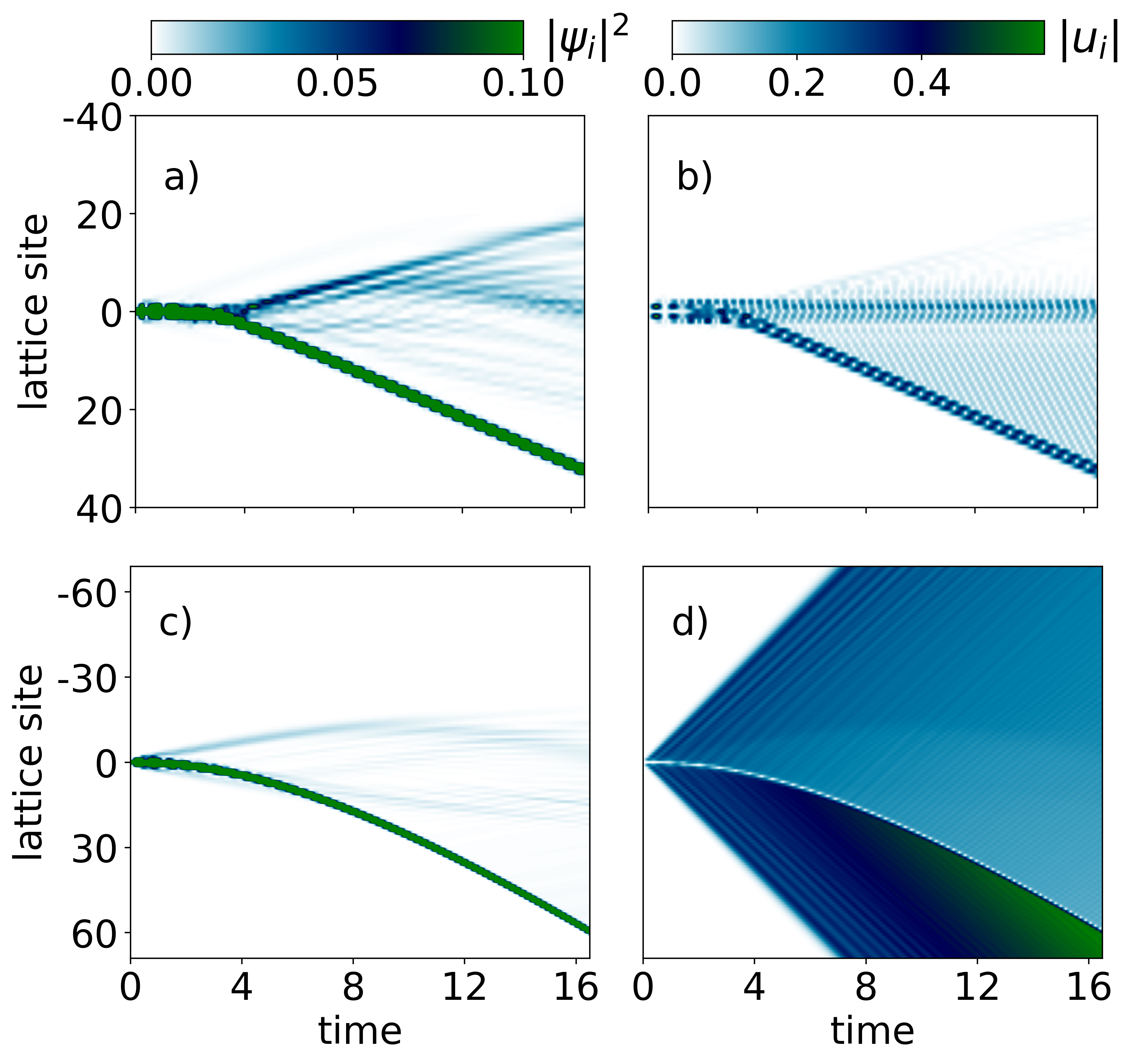}
    \caption{Transport of an excitation coupled to phonons under an external force $F = 0.2$. Panels a) and c) present color-encoded time-evolution of the excitation density $|\psi_i|^2$, while panels b) and d) present color-encoded time-evolution of the phonon field $|u_i|$. The top row corresponds to optical phonons ($g_W = 16$, $g_J = 7$), while the bottom row corresponds to acoustic phonons ($g_W = 17, g_J = 10$).}
    \label{fig:fig6}
\end{figure}

\subsection{Robustness of coherent transport against disorder}
\label{sec:polaron_dynamics_disorder}

\begin{figure}[t!]
    \centering
\includegraphics[scale=0.33]{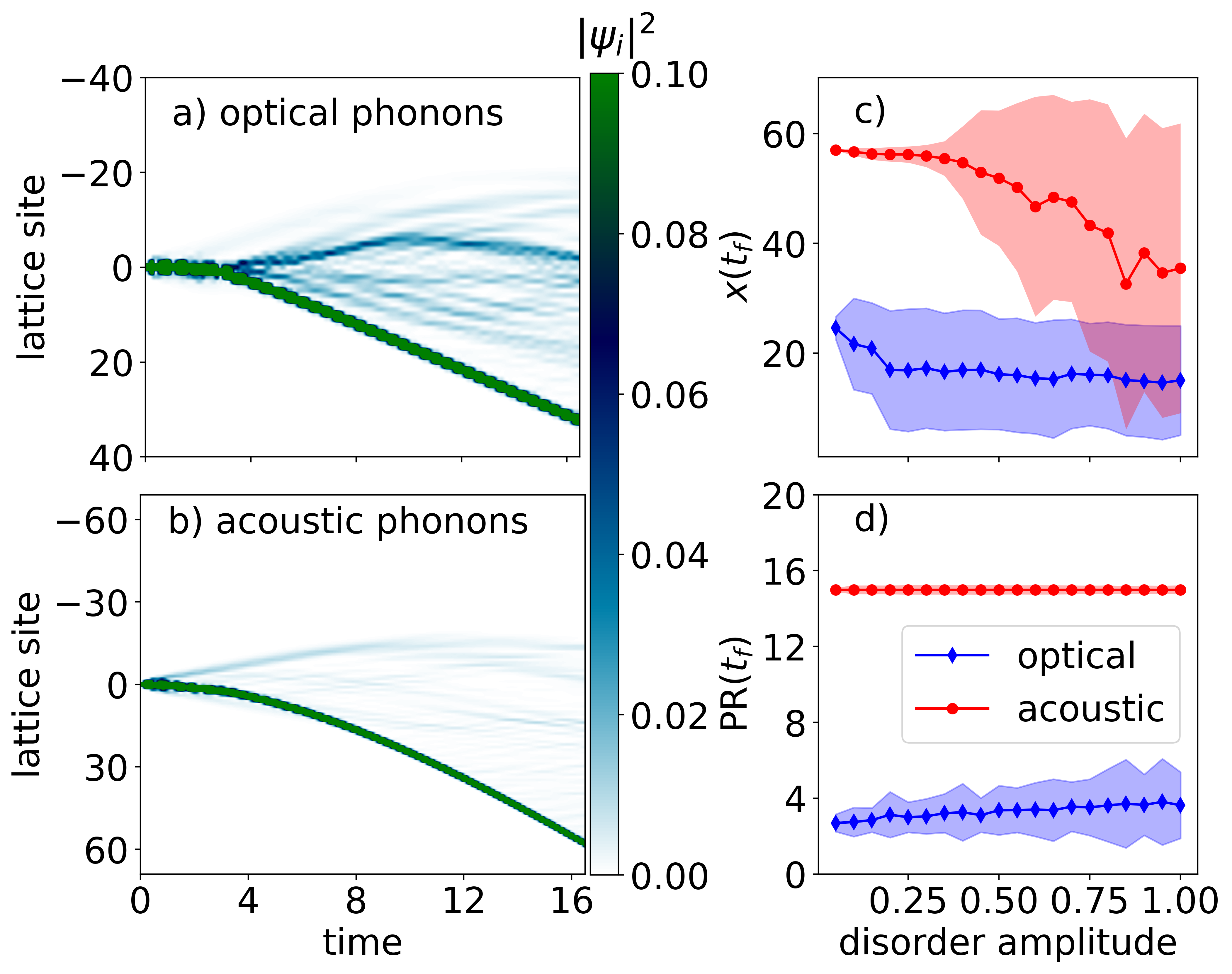}
    \caption{Robustness of non-dispersive moving solutions against on-site disorder, $\epsilon_j~\in~[-W/2,W/2] $. Panels (a) and (b) display  color encoded time evolution of the excitation density, $|\psi_i|^2$, for optical ($g_W = 16$, $g_J = 7$) and acoustic ($g_W = 17$, $g_J = 10$) phonons, respectively, disorder strength $W = 0.6$. Panels (c) and (d) show the center of mass position, $x(t_f)$, and the participation ratio,  $\textrm{PR}(t_f)$,  at the final evolution time  $t_f$, plotted as functions of the disorder amplitude $W$ (top lines (circles) correspond to acoustic phonons, while bottom lines (diamonds) to optical phonons). Although the center of mass positions decreases, a substantial macroscopic transport remains visible.    The remaining parameters used for this analysis are $F=0.2$, $m_{\rm eff}=0.5$, $\omega_{\rm eff} = 10$, $J_0 = 1$.}
    \label{fig:fig7}
\end{figure}

In this paragraph, we focus on the parameters regime, where a well-localized excitation can be transported over a long distance. Namely, after identifying stable polaron regimes, we proceed to apply a constant force to investigate the propagation of non-spreading solutions. 

For this analysis, we fix $F=0.2$, $m_{\rm eff}=0.5$ and select the coupling strengths within the range $g_J \in [4,16]$ and $g_W \in [8,20]$. These regions are indicated by a dashed square in the bottom panels of Fig.~\ref{fig:fig4}. The results are presented in Fig.~\ref{fig:fig5}. The top row of Fig.~\ref{fig:fig5} illustrates the participation ratio, PR$(t_f)$, for both optical [panel (a)] and acoustic [panel (b)] phonons. In both panels, we observe a shift in the boundary between extended and localized states due to the presence of the applied force. However, the prevalence of dark blue colors, indicating localized regimes, remains evident. The bottom row of Fig.~\ref{fig:fig5} displays $\xi(t_f)$, as given by Eq.~\eqref{eq:xi}. This quantity serves as a measure for selecting well-localized solutions propagating in a single direction. We observe stable transport islands of such solutions, indicated by warm colors. Panel (c) corresponds to optical phonons, while panel (d) corresponds to acoustic phonons.

 In Fig.\ref{fig:fig6}, we present an example of transportation of an excitation coupled to phonons under an external force without disorder potential $W=0$. The left column presents the time-evolution of excitation density $|\psi_i|^2$, while the right column presents the evolution of the classical phonon field $|u_i|$.
Next, in Fig.\ref{fig:fig7}, we examine the robustness of the non-dispersive moving solutions against on-site disorder, $\epsilon_j$, as in Eq.\eqref{eq:Hamiltonian_final}. The disorder is introduced by assuming $\epsilon_j$ to be a pseudorandom variable drawn from a uniform distribution in $[-W/2,W/2]$. 
Panels (a) and (b) depict the time propagation of excitations for optical and acoustic phonons, respectively, $W = 0.6$. Panel Fig.~\ref{fig:fig7}(c) illustrates the center of mass position, while Fig.~\ref{fig:fig7}(d) presents the participation ratio evaluated at the final evolution time, plotted as functions of the disorder amplitude $W$. The results are averaged over 200 independent realizations of disorder. Notably, the participation ratio for both acoustic and optical phonons remains relatively constant, providing evidence for the robustness of the polaron self-trapping mechanism, while the center of mass positions takes place on a significant distance.

\section{Summary and conclusions}

In summary, we propose a quantum simulator with Rydbeg-dressed atom arrays for SSH-Fr\"olich Hamiltonian allowing studies of polaron formation and dynamics. The interplay between two competing excitation-phonon coupling terms in the model results in a rich dynamical behavior, which we comprehensively analyze. In particular, our findings reveal the presence of asymmetry in Bloch oscillations allowing coherent transport  of a well-localized excitation over long distances.
Moreover, we compare the behavior of excitations coupled to either acoustic or optical phonons and indicate similar qualitative behavior. Finally, we demonstrate the robustness of phonon-assisted coherent transport to the on-site random potential.

Our analysis is restricted to weak lattice distortions related to a small number of phonons per lattice site, however, the proposed quantum simulator allows the studies of the excitation dynamics in strong distortion limit, as well as studies of a plethora of different scenarios, such as bi- and many-polaron dynamics, and investigation of the quantum boomerang effect \cite{PhysRevA.99.023629, PhysRevB.107.094204,PhysRevX.12.011035} affected by the presence of phonons, both in a single-particle and many-body scenario. We believe, that our work  opens up new avenues for research in Rydberg-based quantum simulators.


\acknowledgments

A.K. acknowledges the support of  the Austrian Science Fund (FWF) within the ESPRIT Programme ESP 171-N under the Quantum Austria Funding Initiative. 
S.K. acknowledges the Netherlands Organisation for Scientific Research (NWO) under Grant No. 680.92.18.05, as well as financial support from the Dutch Ministry of Economic Affairs and Climate Policy (EZK), as part of the Quantum Delta NL program.
ICFO group acknowledges support from: ERC AdG NOQIA; MICIN/AEI (PGC2018-0910.13039/501100011033, CEX2019-000910-S/10.13039/501100011033, Plan National FIDEUA PID2019-106901GB-I00, FPI; MICIIN with funding from European Union NextGenerationEU (PRTR-C17.I1): QUANTERA MAQS PCI2019-111828-2); MCIN/AEI/10.13039/501100011033 and by the “European Union NextGeneration EU/PRTR" QUANTERA DYNAMITE PCI2022-132919 (QuantERA II Programme co-funded by European Union’s Horizon 2020 programme under Grant Agreement No 101017733), Proyectos de I+D+I “Retos Colaboración” QUSPIN RTC2019-007196-7); Fundació Cellex; Fundació Mir-Puig; Generalitat de Catalunya (European Social Fund FEDER and CERCA program, AGAUR Grant No. 2021 SGR 01452, QuantumCAT \ U16-011424, co-funded by ERDF Operational Program of Catalonia 2014-2020); Barcelona Supercomputing Center MareNostrum (FI-2023-1-0013); EU Quantum Flagship (PASQuanS2.1, 101113690); EU Horizon 2020 FET-OPEN OPTOlogic (Grant No 899794); EU Horizon Europe Program (Grant Agreement 101080086 — NeQST), National Science Centre, Poland (Symfonia Grant No. 2016/20/W/ST4/00314); ICFO Internal “QuantumGaudi” project; European Union’s Horizon 2020 research and innovation program under the Marie-Skłodowska-Curie grant agreement No 101029393 (STREDCH) and No 847648 (“La Caixa” Junior Leaders fellowships ID100010434: LCF/BQ/PI19/11690013, LCF/BQ/PI20/11760031, LCF/BQ/PR20/11770012, LCF/BQ/PR21/11840013). Views and opinions expressed are, however, those of the author(s) only and do not necessarily reflect those of the European Union, European Commission, European Climate, Infrastructure and Environment Executive Agency (CINEA), nor any other granting authority. Neither the European Union nor any granting authority can be held responsible for them.

The work J.Z. was funded by the National Science Centre, Poland under the OPUS call within the WEAVE programme
2021/43/I/ST3/01142 as well as  via project 2021/03/Y/ST2/00186 within the QuantERA II Programme that has received funding from the European Union Horizon 2020 research and innovation programme under Grant agreement No 101017733. A
partial support by the Strategic Programme Excellence Initiative  within Priority Research Area (DigiWorld) at Jagiellonian University is acknowledged.
M.P. acknowledges the support of the Polish National Agency for Academic Exchange, the Bekker programme no:
PPN/BEK/2020/1/00317.
Views and opinions expressed are, however, those of the author(s) only and do not necessarily reflect those of the European Union, European Commission, European Climate, Infrastructure and Environment Executive Agency (CINEA), nor any other granting authority. Neither the European Union nor any granting authority can be held responsible for them.

\bibliography{bibliography_polaron}

\end{document}